\documentstyle[12pt]{article}
\newlength{\savetextheight}\savetextheight \textheight
\newlength{\savetopmargin}\savetopmargin \topmargin
\newlength{\saveheadheight}\saveheadheight \headheight
\newlength{\saveheadsep}\saveheadsep \headsep
\newlength{\savetextwidth}\savetextwidth \textwidth
\newlength{\saveoddsidemargin}\saveoddsidemargin \oddsidemargin
\newlength{\saveevensidemargin}\saveevensidemargin \evensidemargin
\newlength{\saveparindent}\saveparindent \parindent

\begin{document}

{\bf\sf
RC 19642  (07/12/94) \\    %?? RC, DATE
Mathematics
}

\vspace{1cm}

\Huge{\bf IBM Research Report}

\vspace*{1cm}

\Large{\bf
An Approximate Fourier Transform Useful in Quantum Factoring
}

\vspace*{1cm}

\normalsize
D. Coppersmith

\vspace*{0.5cm}

IBM Research Division \\
T.J. Watson Research Center \\
Yorktown Heights, New York

\vfill

\scriptsize\sf{\bf\sf LIMITED DISTRIBUTION NOTICE}
 \\
 \\
All rights reserved.
\\
\\

%\parbox{19mm}{\ibmlogo20 IBM}
\parbox{90mm}{\bf\small IBM Research Division \\
Almaden $\cdot$ T.J. Watson $\cdot$ Tokyo $\cdot$ Zurich}

\clearpage
\normalsize
%
%% restores of page attributes added -- cbs 5/10/96
%%
\textheight \savetextheight
\topmargin \savetopmargin
\headheight \saveheadheight
\headsep \saveheadsep
\textwidth \savetextwidth
\oddsidemargin \saveoddsidemargin
\evensidemargin \saveevensidemargin
\parindent \saveparindent
\setcounter{page}{1}
% put in the document but lose "docsytle" and "begin doc"

\def\lemskip{{\vskip 0.4 truecm}}
\def\nq{{\mbox{$\frac{n}{4}$}}}
\def\nh{{\mbox{$\frac{n}{2}$}}}

\newcommand{\Z}{{\bf Z}}
\newcommand{\Q}{{\bf Q}}
\newcommand{\oA}{\overline{A}}
\newcommand{\sA}{\overline{A}^{*}}
\newtheorem{lemma}{Lemma}

\title{An Approximate Fourier Transform Useful in Quantum Factoring}
\author{Don Coppersmith}
\date{June, 1994}
\maketitle

%\begin{document}

{\bf Abstract.}
We define an approximate version of the Fourier transform
on $2^L$ elements,
which is computationally attractive in a certain setting,
and which may find application to the problem of factoring integers
with a quantum computer as is currently under investigation
by Peter Shor. [SHO]

{\bf Fourier Transform}

{\bf Notation:}
Let $L$ be a positive integer.
Let $a,c$ be $L$-bit integers.
The binary representations of $a, c$ are
$$a = \Sigma_{i=0}^{L-1} a_i 2^i ,
  c = \Sigma_{i=0}^{L-1} c_i 2^i .$$
Define the $L$-bit integer $b$ as the reversal of $c,$
$$b = \Sigma b_i 2^i = \Sigma c_{L-1-i} 2^i ,$$
so that $b_i = c_{L-1-i} .$
Let $X, Y$ be arrays of size $2^L$ indexed by $a$ or
$c.$
Let $\omega = \omega_{( 2^L )} = \exp \left( 2 \pi i / 2^L \right)$
be the standard $2^L$ root of unity.

The ordinary Fourier transform is defined as
$$Y_c = \frac{1}{\sqrt { 2^L }}
\Sigma_a X_a \omega ^ { a c }
= \frac{1}{\sqrt { 2^L }}
\Sigma_a X_a \exp \left( \frac {2 \pi i }{ 2^L } a c \right) $$
In terms of the binary representations,
$$Y_c = \frac {1} { \sqrt { 2^L }} \Sigma_a X_a
\exp \left( \frac{ 2 \pi i }{ 2^L }
\Sigma_{j,k=0}^{L-1} a_j c_k 2^{j+k} \right) $$
Whenever $j+k \ge L ,$ we have $\omega^{(2^{j+k})} =1,$
so that we can drop those terms from consideration:
$$(FFT) ~~~~~~~~~~~~~~~~~
Y_c = \frac{1}{\sqrt { 2^L}} \Sigma_a X_a
\exp \left( \frac{ 2 \pi i }{ 2^L }
\Sigma_{0 \le j,k \le L-1, j+k \le L-1 }
a_j c_k 2^{j+k} \right)  $$
\newpage
Notice that all computations are in the field
$\Q ( \omega_{( 2^L )} ).$

{\bf Hadamard Transform}

The Hadamard transform looks like a Fourier transform defined
over $\Z_2^L .$
It suits my purposes, pedagogically, to reverse the indexing
on the output of the Hadamard transform, and get the transform
$$
Y_c = \frac{1}{\sqrt {2^L}} \Sigma_a X_a
(-1) ^ {( \Sigma_j a_j  c_{L-1-j} ) }$$
(Normally the exponent would be
$\Sigma_j a_j c_j ,$
but we reverse the indexing to bring out the similarity with
the ordinary FFT.)
Rewrite this as
$$ Y_c = \frac{1}{\sqrt {2^L}}\Sigma_a X_a
\exp \left( \frac{ 2 \pi i}{ 2^L }
\Sigma_{j=0}^{L-1} a_j c_{L-1-j} 2^{L-1} \right)$$
$$ = \frac{1} {\sqrt {2^L}} \Sigma_a X_a
\exp \left( \frac{2 \pi i}{2^L}
\Sigma_{0 \le j,k \le L-1; j+k=L-1}
a_j c_k 2^{L-1} \right) , $$
noting that $\exp ( 2 \pi i 2^{L-1} /2^L) = -1 .$
Since the sum is restricted to those values of $j,k$ satisfying
$j+k=L-1,$ we can replace
$2^{L-1}$ by $2^{j+k}$ and obtain
$$(HT) ~~~~~~~~~~~~~~~~~
Y_c = \frac{1}{\sqrt { 2^L }}\Sigma_a X_a
\exp \left( \frac{2 \pi i }{ 2^L}
\Sigma_{0 \le j,k \le L-1; j+k=L-1}
a_j c_k 2^{j+k} \right) $$

{\bf Approximate Fourier Transform}

Comparing the two formulas (FFT) and (HT),
we find that the only difference is in the limits on $j+k:$
in (FFT) the range is $0 \le j+k \le L-1,$
while in (HT) the range is $L-1 \le j+k \le L-1 .$

This leads us to define an Approximate Fourier Transform
(AFFT), parameterized by an integer $m:$
$$(AFFT_m) ~~~~~~~~~~~~~~~~~
Y_c
= \frac{1}{\sqrt {2^L}} \Sigma_a X_a
\exp \left( \frac{ 2 \pi i}{ 2^L }
\Sigma_{0 \le j,k \le L-1; L-m \le j+k \le L-1}
a_j c_k 2^{j+k} \right) $$
When $m=1$ this is the Hadamard transform (suitably indexed);
when $m=L$ it becomes the ordinary Fourier transform.

Since $j+k \ge L-m,$ the argument of ``exp''
is some multiple of $2 \pi i 2^{L-m} / 2^L = 2 \pi i / 2^m ,$
so that AFFT is defined over $\Q ( \omega_{( 2^m )} ) .$

The argument of ``exp'' in AFFT differs from that of
FFT by
$$\frac{ 2 \pi i }{ 2^L} \Sigma_{ j+k < L-m }
 a_j c_k 2^{j+k}  .$$
The magnitude of this difference is bounded by
$$\frac{ 2 \pi }{ 2^L } L 2^{L-m} = 2 \pi L 2^{-m} .$$
If $L=500$ and $m=20,$ this bound is
about 3/1000.
So the matrix entries of AFFT differ from those of FFT
by a multiplicative factor of $\exp ( i \epsilon ) $ where
$| \epsilon | \le 2 \pi L 2^{-m} = 3/1000 .$
Thus if AFFT is used in place of FFT in Shor's factoring
work [SHO],
it leads to an overall error of a fraction of a degree in
each phase angle, and less than one percent decrease in the magnitude
of the probability of each desirable final state.

{\bf Calculating the AFFT}

Start with the description of the Fast Fourier Transform
as taken from [KNU, page 291, section 4.3.3].
I have replaced $A,t,s,k$ by $X,a,b,L,$ respectively,
and numbered the passes from $L-1$ down to 0,
to correspond to the bit being manipulated.

* Initialization.
Let $X^{[L]} ( a_{L-1} , ... , a_0 )
= X_a ,$ where $a = ( a_{L-1} ... a_0 )_2$
(the binary representation).

* Pass $J, J=L-1, L-2, ... , 1, 0.$ (Numbered downwards!)  Set

* $X^{[J]} ( b_{L-1}, b_{L-2} , ..., b_J , a_{J-1} ,
... , a_0 ) :=$

~~~~~~~~$  X^{[J+1]} ( b_{L-1} , ... , b_{J+1} , 0 ,
a_{J-1} , ... , a_0 ) + $

~~~~~~~~$ \omega^ { (b_J b_{J+1} ... b_{L-1} 0 ... 0 )_2 }
\times X^{[J+1]} ( b_{L-1} , ... , b_{J+1} , 1 , a_{J-1} ,
... , a_0 ) $

We wish to compute the FFT quantum mechanically.
At the outset, $X^{[L]} ( a_{L-1} , ... , a_0 )$
represents the amplitude of the state where $L$ electrons
have spins $a_{L-1} , ... , a_0 ,$ respectively,
with ``1'' representing ``up''
and ``0'' representing ``down''.
Each succeeding $X^{[J]} ( b_{L-1}  , ... , a_0 ) $
represents the amplitude of the state of these same $L$
electrons.
The transform is performed by a sequence of two-electron interactions.

On Pass $J,$ multiply the amplitudes
$X^{[J+1]} ( b_{L-1} , ... , b_{J+1} , 1 , a_{J-1} ,
... , a_0 ) $
(with a 1 in position $J$) by the phase shift
$\omega^ { (0 b_{J+1} ... b_{L-1} 0...0 )_2 } .$
This correspond to the following two-bit operations.
For each $K,$ $J+1 \le K \le L-1 ,$
use an interaction between electrons $J$ and $K$ to
multiply the amplitude of those states with a 1 in both positions
$J$ and $K$ by the factor
$$\omega ^ {( 2 ^ {L-1-K+J} )} .$$
Call this transformation $Q_{JK} .$

Then apply the unitary transformation
$$\frac{1}{\sqrt 2}
\left(
\begin{array}{cc}
1 & 1 \\
1 & -1 \\
\end{array}
\right) $$
to the electron $J$.
Call this transformation $P_J .$
So for $L=3, J=1,$ the only value of $K$ is $K=2,$
and we have
$$\frac{1}{\sqrt 2}
\left[ \begin{array}{cccccccc}
1 & 0 & {\omega ^0} & 0 & 0 & 0 & 0 & 0 \\
0 & 1 & 0 & {\omega ^0} & 0 & 0 & 0 & 0 \\
1 & 0 & {\omega ^4} & 0 & 0 & 0 & 0 & 0 \\
0 & 1 & 0 & {\omega ^4} & 0 & 0 & 0 & 0 \\
0 & 0 & 0 & 0 & 1 & 0 & {\omega ^2} & 0 \\
0 & 0 & 0 & 0 & 0 & 1 & 0 & {\omega ^2} \\
0 & 0 & 0 & 0 & 1 & 0 & {\omega ^6} & 0 \\
0 & 0 & 0 & 0 & 0 & 1 & 0 & {\omega ^6}
\end{array} \right]
= P_1 Q_{12} = $$
$$= \frac{1}{\sqrt 2}
\left[ \begin{array}{cccccccc}
1 & 0 & 1 & 0 & 0 & 0 & 0 & 0 \\
0 & 1 & 0 & 1 & 0 & 0 & 0 & 0 \\
1 & 0 & -1 & 0 & 0 & 0 & 0 & 0 \\
0 & 1 & 0 & -1 & 0 & 0 & 0 & 0 \\
0 & 0 & 0 & 0 & 1 & 0 & 1 & 0 \\
0 & 0 & 0 & 0 & 0 & 1 & 0 & 1 \\
0 & 0 & 0 & 0 & 1 & 0 & -1 & 0 \\
0 & 0 & 0 & 0 & 0 & 1 & 0 & -1
\end{array} \right] \times
\left[ \begin{array}{cccccccc}
1 & 0 & 0 & 0 & 0 & 0 & 0 & 0 \\
0 & 1 & 0 & 0 & 0 & 0 & 0 & 0 \\
0 & 0 & 1 & 0 & 0 & 0 & 0 & 0 \\
0 & 0 & 0 & 1 & 0 & 0 & 0 & 0 \\
0 & 0 & 0 & 0 & 1 & 0 & 0 & 0 \\
0 & 0 & 0 & 0 & 0 & 1 & 0 & 0 \\
0 & 0 & 0 & 0 & 0 & 0 & {\omega ^2} & 0 \\
0 & 0 & 0 & 0 & 0 & 0 & 0 & {\omega ^2}
\end{array} \right] $$
In general one would have $L-1-J$ of the two-bit interactions $Q_{JK}$
on pass $J$, corresponding to different values of $K$.
The entire 3-spin FFT is depicted in the Appendix.

So the FFT matrix is expressed as a product of unitary matrices.
For example FFT on 4 electrons is
$$P_0 Q_{01} Q_{02} Q_{03} P_1 Q_{12} Q_{13} P_2 Q_{23} P_3 .$$
If there are $L$ electrons then there are $L$ matrices
$P_J$ and $L(L-1)/2$ matrices $Q_{JK}$.

For our approximate AFFT, we simply delete those
matrices $Q_{JK}$ with $K \ge J+ m .$
So the AFFT is again unitary, and easily computed with
one-bit and two-bit operators.
It requires about $L m$ two-bit operations.

{\bf Quantum computation}

Shor [SHO, page 12] suggests first developing a state
$$\frac{1}{\sqrt q} \Sigma_{a=0}^{q-1} | a > $$
where $q \approx 5 n^2$ is a product of small prime powers,
which will enable him to do a mixed-radix FFT later.
By contrast, we suggest setting $q = 2^L \approx 5 n^2 .$
Second, he computes $x^a (mod~n) ,$ where $x,n$
are integers computed classically,
so that the state becomes
$$\frac{1}{\sqrt q} \Sigma_{a=0}^{q-1} | a , x^a > $$
Then he does the Fourier transform, sending $a$ to $c$
with amplitude $\frac{1}{\sqrt q} \exp ( 2 \pi i a c / q ) .$
This leaves the machine in state
$$\frac{1}{q} \Sigma_{a,c=0}^{q-1} \exp ( 2 \pi i a c / q )
| c , x^a > .$$

We see that the radix-$2^L$ Fourier transform
is directly implementable as $L^2$ 2-spin interactions,
as opposed to the $L^3$ operations required by Shor.

We can improve still further, by doing
our approximate Fourier transform instead of the
Fourier transform.
Notice that on Pass $J$ of AFFT computation,
we use interactions between bits $J$ and $K$,
$J < K < J+m $.
So bit $K$ of the output index, $b_K $, does not
participate in any interaction after pass $J=K-m$.
(Remember we are numbering backwards, so pass $K-m$ is $m$
passes later than pass $K$.)
Similarly, bit $K$ of the input index, $a_K$,
does not enter into the computation until pass $J=K$.

So we propose rearranging the computation in the following way.

* Start with $y=1$ in an $L$-bit quantum register

~~~~~~~~~~where you will compute $x^a$.

* For each $J=L-1,L-2,...,2,1,0:$

~~~* Place the electron $J$ in state
$$\frac{1}{\sqrt 2} ( | 0 > + | 1 > ) $$
~~~~~~~~~~corresponding to the two possible values of $a_J .$

~~~* Compute $$y := y ( x ^ { 2^J } )^ {( a_J )} (mod~n) $$
~~~~~~~~~~reversibly, in the register allocated for $y$.

~~~* For $K=J+1,J+2,..., \min (J+m-1,L-1),$ apply operation $Q_{JK}$.

~~~* Apply operation $P_J$.

~~~* If $J \le L-m$, measure bit $b_{J+m-1} = c_{L-J-m}$ from the output

~~~~~~~~~~of pass $J$ of the AFFT computation.

~~~~~~~~~~(It will not enter any more interactions.)

* End (For each $J=L-1, L-2,...,2,1,0$).

* Measure the remaining bits $b_{m-2} , ... , b_0$.

* End algorithm

A possible advantage of this arrangement is that the electron
in position $K$ need only maintain coherence for $m$
passes of the computation, although the rest of the system
still has to maintain coherence for a longer time, so this
advantage might be less than it appears at first blush.

A definite advantage is in the computational complexity.
Shor's proposal, using a mixed-radix Fourier Transform
with $q \approx 5n^2$ the product of small prime powers,
appears to require about $(\log n)^3$ elementary operations
(spin-spin interactions).  The radix-$2^L$ FFT requires
only $(\log n)^2$ elementary operations.
The AFFT requires only $(\log n)(\log \log n + \log 1/\epsilon )$
operations, where a final precision of $\epsilon$ is required.
So the Fourier transform is no longer the bottleneck of the
computation.

{\bf Parallel implementation}

Several steps of the AFFT can be parallelized in the
quantum implementation; this might further speed up the
computation time, and increase the likelihood of the state
remaining coherent until the computation is done.

We use, for an example,
the FFT on 5 electrons, with operations proceeding right to left:
$$FFT = P_0 Q_{01} Q_{02} Q_{03} Q_{04}
        P_1 Q_{12} Q_{13} Q_{14}
        P_2 Q_{23} Q_{24}
        P_3 Q_{34}
        P_4$$
We can interchange the order of any two operations which do not
involve any of the same electrons; alternatively, we can do
such operations in parallel.
At time step $K = 8,7,...,1,0,$
let us perform $P_I$ if
$I+I=K$, and $Q_{IJ}$ if $I+J=K$.
Steps that are performed in parallel are displayed within
square brackets, vertically aligned, and again proceeding
right to left:
$$FFT = P_0 Q_{01}
\left[ \begin{array}{c} P_1 \\ Q_{02} \end{array} \right]
\left[ \begin{array}{c} Q_{12} \\ Q_{03} \end{array} \right]
\left[ \begin{array}{c} P_2 \\ Q_{13} \\ Q_{04} \end{array} \right]
\left[ \begin{array}{c} Q_{23} \\ Q_{14} \end{array} \right]
\left[ \begin{array}{c} P_3 \\ Q_{24} \end{array} \right]
Q_{34} P_4 $$

We used 9 time steps here; for an $L$-electron system
we will use $2L-1$ time steps.

This parallel implementation looks a lot like
``systolic arrays,'' [MC, chapter 8, section 8.3],
and suggests directions for physical implementation.

{\bf References}

[KNU]
Donald E. Knuth, volume 2.
The Art of Computer Programming,
Volume 2: Seminumerical Algorithms.
(Addison-Wesley, Reading, MA, 2nd ed., 1981)

[MC]
Carver Mead and Lynn Conway,
Introduction to VLSI Systems.
(Addison-Wesley, Reading, MA, 1980)

[SHO]
Peter W. Shor,
``Algorithms for Quantum Computation: Discrete Log and
Factoring,''
manuscript, 1994. Proceedings of FOCS 1994.

{\bf Appendix}

We write out in full the FFT on 3 electrons.
Note that the rows are numbered in bit-reversed order
(04261537), corresponding to the index of $b$ rather than $c$.
$$ FFT = P_0 Q_{01} Q_{02} P_1 Q_{12} P_2 = $$
$$ =
\frac {1}{\sqrt 2}
\left[ \begin{array}{cccccccc}
1 & 1 & 0 & 0 & 0 & 0 & 0 & 0 \\
1 & -1 & 0 & 0 & 0 & 0 & 0 & 0 \\
0 & 0 & 1 & 1 & 0 & 0 & 0 & 0  \\
0 & 0 & 1 & -1 & 0 & 0 & 0 & 0 \\
0 & 0 & 0 & 0 & 1 & 1 & 0 & 0  \\
0 & 0 & 0 & 0 & 1 & -1 & 0 & 0 \\
0 & 0 & 0 & 0 & 0 & 0 & 1 & 1  \\
0 & 0 & 0 & 0 & 0 & 0 & 1 & -1 \\
 & & & P_0 & & & &
\end{array} \right]
          \times
\left[ \begin{array}{cccccccc}
1 & 0 & 0 & 0 & 0 & 0 & 0 & 0 \\
0 & 1 & 0 & 0 & 0 & 0 & 0 & 0 \\
0 & 0 & 1 & 0 & 0 & 0 & 0 & 0 \\
0 & 0 & 0 & \omega^2 & 0 & 0 & 0 & 0 \\
0 & 0 & 0 & 0 & 1 & 0 & 0 & 0 \\
0 & 0 & 0 & 0 & 0 & 1 & 0 & 0 \\
0 & 0 & 0 & 0 & 0 & 0 & 1 & 0 \\
0 & 0 & 0 & 0 & 0 & 0 & 0 & \omega^2 \\
 & & & Q_{01} & & & &
\end{array} \right]
     \times $$
$$ \times
\left[ \begin{array}{cccccccc}
1 & 0 & 0 & 0 & 0 & 0 & 0 & 0 \\
0 & 1 & 0 & 0 & 0 & 0 & 0 & 0 \\
0 & 0 & 1 & 0 & 0 & 0 & 0 & 0 \\
0 & 0 & 0 & 1 & 0 & 0 & 0 & 0 \\
0 & 0 & 0 & 0 & 1 & 0 & 0 & 0 \\
0 & 0 & 0 & 0 & 0 & \omega & 0 & 0 \\
0 & 0 & 0 & 0 & 0 & 0 & 1 & 0 \\
0 & 0 & 0 & 0 & 0 & 0 & 0 & \omega \\
 & & & Q_{02} & & & &
\end{array} \right]
\times \frac{1}{\sqrt 2}
\left[ \begin{array}{cccccccc}
1 & 0 & 1 & 0 & 0 & 0 & 0 & 0 \\
0 & 1 & 0 & 1 & 0 & 0 & 0 & 0 \\
1 & 0 & -1 & 0 & 0 & 0 & 0 & 0 \\
0 & 1 & 0 & -1 & 0 & 0 & 0 & 0 \\
0 & 0 & 0 & 0 & 1 & 0 & 1 & 0 \\
0 & 0 & 0 & 0 & 0 & 1 & 0 & 1 \\
0 & 0 & 0 & 0 & 1 & 0 & -1 & 0 \\
0 & 0 & 0 & 0 & 0 & 1 & 0 & -1 \\
 & & & P_1 & & & &
\end{array} \right]
 \times $$
$$ \times
\left[ \begin{array}{cccccccc}
1 & 0 & 0 & 0 & 0 & 0 & 0 & 0 \\
0 & 1 & 0 & 0 & 0 & 0 & 0 & 0 \\
0 & 0 & 1 & 0 & 0 & 0 & 0 & 0 \\
0 & 0 & 0 & 1 & 0 & 0 & 0 & 0 \\
0 & 0 & 0 & 0 & 1 & 0 & 0 & 0 \\
0 & 0 & 0 & 0 & 0 & 1 & 0 & 0 \\
0 & 0 & 0 & 0 & 0 & 0 & \omega^2 & 0 \\
0 & 0 & 0 & 0 & 0 & 0 & 0 & \omega^2 \\
 & & & Q_{12} & & & &
\end{array} \right]
\times \frac{1}{\sqrt 2}
\left[ \begin{array}{cccccccc}
1 & 0 & 0 & 0 & 1 & 0 & 0 & 0 \\
0 & 1 & 0 & 0 & 0 & 1 & 0 & 0 \\
0 & 0 & 1 & 0 & 0 & 0 & 1 & 0 \\
0 & 0 & 0 & 1 & 0 & 0 & 0 & 1 \\
1 & 0 & 0 & 0 & -1 & 0 & 0 & 0 \\
0 & 1 & 0 & 0 & 0 & -1 & 0 & 0 \\
0 & 0 & 1 & 0 & 0 & 0 & -1 & 0 \\
0 & 0 & 0 & 1 & 0 & 0 & 0 & -1 \\
 & & & P_2 & & & &
\end{array} \right] $$
$$ = \frac{1}{\sqrt 8}
\left[ \begin{array}{cccccccc}
\omega^0 & \omega^0 & \omega^0 & \omega^0 &
\omega^0 & \omega^0 & \omega^0 & \omega^0 \\
\omega^0 & \omega^4 & \omega^0 & \omega^4 &
\omega^0 & \omega^4 & \omega^0 & \omega^4 \\
\omega^0 & \omega^2 & \omega^4 & \omega^6 &
\omega^0 & \omega^2 & \omega^4 & \omega^6 \\
\omega^0 & \omega^6 & \omega^4 & \omega^2 &
\omega^0 & \omega^6 & \omega^4 & \omega^2 \\
\omega^0 & \omega^1 & \omega^2 & \omega^3 &
\omega^4 & \omega^5 & \omega^6 & \omega^7 \\
\omega^0 & \omega^5 & \omega^2 & \omega^7 &
\omega^4 & \omega^1 & \omega^6 & \omega^3 \\
\omega^0 & \omega^3 & \omega^6 & \omega^1 &
\omega^4 & \omega^7 & \omega^2 & \omega^5 \\
\omega^0 & \omega^7 & \omega^6 & \omega^5 &
\omega^4 & \omega^3 & \omega^2 & \omega^1
\end{array} \right] $$
\end{document}